\documentclass[manuscript]{aastex}

\slugcomment{Accepted by Astrophysical Journal}

\shorttitle{Solar Flare Iron Abundance}
\shortauthors{Phillips and Dennis}

\begin{document}


\title{The Solar Flare Iron Abundance}


\author{K. J. H. Phillips\altaffilmark{1}}
\affil{Mullard Space Science Laboratory, University College London, Holmbury St Mary, Dorking RH6 5NT, UK}
\email{kjhp@mssl.ucl.ac.uk}

\author{B. R. Dennis\altaffilmark{2}}
\affil{NASA Goddard Space Flight Center, Greenbelt, MD 20771}
\email{Brian.R.Dennis@nasa.gov}

\begin{abstract}
The abundance of iron is measured from emission line complexes at 6.65~keV (Fe line) and 8~keV (Fe/Ni line) in {\em RHESSI} X-ray spectra during solar flares. Spectra during long-duration flares with steady declines were selected, with an isothermal assumption and improved data analysis methods over previous work. Two spectral fitting models give comparable results, viz. an iron abundance that is lower than previous coronal values but higher than photospheric values. In the preferred method, the estimated Fe abundance is $A({\rm Fe}) = 7.91 \pm 0.10$ (on a logarithmic scale, with $A({\rm H}) = 12$), or $2.6 \pm 0.6$ times the photospheric Fe abundance. Our estimate is based on a detailed analysis of 1,898 spectra taken during 20 flares. No variation from flare to flare is indicated. This argues for a fractionation mechanism  similar to quiet-Sun plasma. The new value of $A({\rm Fe})$ has important implications for radiation loss curves, which are estimated.
\end{abstract}


% Section 1
\section{INTRODUCTION}

The solar abundance of iron remains an important parameter and topic in solar physics. Iron is the most abundant of all elements with $Z > 14$, and is a large contributor to the radiation loss at coronal temperatures. Recent determinations of the photospheric abundance give $A({\rm Fe}) = 7.50\pm 0.04$ \citep{asp09} and  $7.52 \pm 0.06$ \citep{caf11} (on a logarithmic scale where $A({\rm H}) = 12$), in near-agreement with the meteoritic abundance, $A({\rm Fe}) = 7.45\pm 0.01$ \citep{lod09}. The iron abundance in the corona has been determined from X-ray or extreme ultraviolet Fe emission lines, formed by collisional excitation. As the excitation rates are a function of electron temperature $T_e$, the thermal structure of the emitting coronal plasma must be modeled for correct interpretation of line fluxes, and ionization fractions and excitation rate coefficients must be known. Examples of the coronal Fe abundance include \cite{par77} ($A({\rm Fe}) = 7.65$), \cite{flu99} (7.65), \cite{whi00} (8.19), and \cite{den08} (7.86), i.e. enhancement factors over the photospheric value of between 1.4 and 4.9. This large range of abundance determinations may indicate time variations in the coronal abundance \citep{syl84} or measurement uncertainties of $\sim 0.5$ in $A({\rm Fe})$.  The Fe abundance from solar energetic particles (SEPs) in the interplanetary medium varies by large factors; a baseline list of abundances for gradual events \citep{rea95} gives $A({\rm Fe}) = 7.93$, or 2.65 times photospheric. Evidence for systematic differences between photospheric and coronal abundances has been cited by \cite{mey85} and \cite{fel92}. According to \cite{fel00} elements with first ionization potential FIP $\lesssim 10$~eV like Fe have coronal abundances enhanced by factors of 4, apart from low-altitude flares and energetic spray-like events for which the coronal and photospheric abundances are equal. Various models have been put forward to explain the fractionation and its dependence on FIP; they generally involve a mechanism that separates ions and neutral atoms in the chromosphere where low-FIP elements are partly ionized but high-FIP elements are neutral. The mechanisms include magnetic fields carrying ions rising into the corona as active regions develop \citep{hen98}, and the presence of a ponderomotive force in the acceleration of Alfv\'en waves \citep{lam09}.

Flare spectra in the photon energy range $\sim 3$~keV to 17~MeV have been obtained from the {\it Reuven Ramaty High Energy Solar Spectroscopic Imager} ({\em RHESSI}) since its launch on 2002 February~5, allowing analysis of thermal spectra (range $\sim 6$--30~keV) with $\sim 1$~keV spectral resolution (FWHM). This range includes thermal continuum emission (free--free and free-bound) and two line complexes \citep{phi04}. The Fe line complex at $\sim 6.65$~keV is made up of \ion{Fe}{25} lines and dielectronic satellites of \ion{Fe}{24} and lower stages, with minor contributions from the \ion{Fe}{26} Lyman-$\alpha$ lines \citep{fel80,tan82,lem84}, emitted over a broad temperature range ($\gtrsim 10$--100~MK). The weaker ``Fe/Ni line" line complex at $\sim 8$ keV consists of higher-excitation lines ($1s^2 - 1snp$, $n \geqslant 3$) of \ion{Fe}{25} and \ion{Fe}{24} satellites, with a few-percent contribution from \ion{Ni}{26}--\ion{Ni}{28} lines. Line-to-continuum ratios can be derived from {\em RHESSI} spectra, and from these the iron abundance estimated using models for the thermal structure (isothermal or simple functions for the temperature distribution). This was done for {\em RHESSI} spectra taken during various phases of 27 flares between 2002 and 2005 \citep{phi06} using fluxes of the Fe line complex at 6.65~keV expressed as the equivalent width of nearby continuum. The observed equivalent widths were found to follow a dependence on $T_e$, derived from the energy dependence of the continuum emission, approximately equal to the theoretical dependence with an Fe abundance somewhat less than \cite{fel00}'s coronal value, $A({\rm Fe}) = 8.10$. In this work, this analysis is carried further. First, we have chosen only spectra during the gradual phases of flares with the {\em RHESSI} thin attenuators in place, and secondly we have used advances in the analysis software enabling {\em RHESSI} spectra to be better interpreted, including the use of the {\sc chianti} code (v.~6) \citep{der97,you03,der09} with latest atomic data instead of the earlier {\sc mekal} code \citep{mew85}.  Also, instrumental effects can be adjusted in the fitting process. By these means, we derive an estimate of the Fe abundance with much reduced uncertainties compared with previous work.

% Section 2
\section{SPECTRAL ANALYSIS AND OBSERVATIONS}

% Subsection 2.1
\subsection{The {\em RHESSI} Spectrometer}

The {\em RHESSI} spectrometer has been described by \cite{smi02} so only an outline is given here. Nine cryogenically cooled hyper-pure germanium detectors make up the spectrometer, each with a 1-cm-thick front segment which senses low-energy X-rays. Modulation collimators in front of each detector result in time-modulated counting rates as the spacecraft rotates, which are unscrambled with imaging software to form images. The X-ray emission is viewed through and partially absorbed by beryllium windows and aluminized Mylar insulation layers. To avoid detector saturation and to reduce pulse pile-up, sets of aluminum attenuators are moved in front of the detectors at increasing X-ray flux levels. Since instrumental effects like pile-up, slight changes in energy calibration at high count rates, and variable background rates are different for each detector, we chose to analyze spectra from the detector with best energy resolution rather than summing counts from multiple detectors to improve statistics. Following our earlier work \citep{phi06}, we selected flare spectra during times of slowly decaying emission in the A1 state (thin attenuators in place) when the flare plasma is most nearly isothermal. Spectral fits were done over a range from $\sim 5.7$~keV (the count rates at lower energies are dominated by K-escape events in the A1 and A3 attenuator states) up to 20--40~keV, depending on the emission at higher energies relative to the background spectrum. Thick-target X-ray continuum from a power-law electron spectrum was included when necessary to achieve an acceptable fit to the data at the higher energies. The energy bins for the spectral fits were those $\frac{1}{3}$~keV-wide bins used by the on-board pulse-height analyzer.

% Subsection 2.2
\subsection{Data Selection}

Twenty flares having a steady decline of X-ray emission were observed by {\em RHESSI} between 2002 and 2005 in its A1 attenuator state during the $\sim 60$-minute solar-viewing part of its orbit. For two long-duration flares (on 2002 July 20/21 and July 26/27), the decline could be followed for up to four orbits. Table~\ref{anal_ints} gives the dates and {\em GOES} classes with time intervals over which analysis of at least the Fe-line complex was possible, one interval per flare except for the 2002 July 20/21 and July 26/27 flares; numbers indicate flares, letters time intervals for the 2002 July 20/21 and 26/27 flares. Also listed are the number of spectra $N_{\rm full}$, and $N_{\rm cont}$ analyzed in each interval for each of the {\sc chianti$\_$full} and {\sc chianti$\_$cont} emission models, and details of the estimated Fe abundance which will be discussed in Section~\ref{Fe_abund_anal}.

% Table 1: List of time intervals and abundance values for RHESSI flares
\begin{deluxetable}{rlccrcrccc}
\rotate
\tabletypesize{\footnotesize} \tablecaption{T{\sc ime} I{\sc ntervals} A{\sc nalyzed during} {\em RHESSI}
F{\sc lares} \label{anal_ints}} \tablewidth{0pt}
\tablehead{\colhead{Flare } & \colhead{Date} & \colhead{{\em RHESSI} Time} &\colhead{{\em GOES}}& & \colhead{{\sc chianti$\_$full}}& Analysis & &\colhead{{\sc chianti$\_$cont}} & Analysis\\
& & \colhead{Range (UT)} & \colhead{Class} &  & \colhead{Mean value} & & & \colhead{$A({\rm Fe})$} & \colhead{$A({\rm Fe})$} \\
& & & & \colhead{$N_{\rm full}$} & \colhead{of $F$} & \colhead{$A({\rm Fe})$} & \colhead{$N_{\rm cont}$} &\colhead{Fe-Line} &\colhead{Fe/Ni-Line} \\
& & & & & & & & \colhead{Complex} & \colhead{Complex} \\

}

\startdata
1..& 2002 Mar 10     & 22:57 -- 23:50 & M2.3&   9 & $0.54 \pm 0.04$ & $7.83 \pm 0.03$ & 186 & $7.91 \pm 0.03$ & $8.02 \pm 0.07$ \\
2..& 2002 Apr 15     & 00:15 -- 00:40 & M3.7&  59 & $0.56 \pm 0.19$ & $7.85 \pm 0.13$ &  88 & $7.86 \pm 0.09$ & $8.02 \pm 0.10$ \\
3..& 2002 May 31     & 00:13 -- 00:57 & M2.4&     & indeterminate$^b$ &               & 149 & $7.93 \pm 0.13$ & $8.04 \pm 0.14$ \\
4 ..& 2002 Jun 1     & 03:52 -- 04:04 & M1.6&  17 & $0.93 \pm 0.55$ & $8.07 \pm 0.20$ &  36 & $8.01 \pm 0.09$ & indeterminate \\
5a..& 2002 Jul 20/21 & 22:29 -- 23:28 & X3.3&  21 & $0.52 \pm 0.07$ & $7.81 \pm 0.06$ &  90 & $7.80 \pm 0.08$ & $7.87 \pm 0.20$ \\
5b..&                & 00:06 -- 01:03 &     & 205 & $0.53 \pm 0.22$ & $7.82 \pm 0.16$ & 172 & $7.89 \pm 0.12$ & $8.07 \pm 0.16$ \\
5c..&                & 01:42 -- 02:40 &     &     & indeterminate   &                 & 172 & $7.87 \pm 0.20$ & indeterminate \\
5d..&                & 03:24 -- 04:16 &     &     & indeterminate   &                 &  86 & $8.03 \pm 0.21$ & indeterminate \\
6a..& 2002 Jul 26/27 & 23:01 -- 00:00 & M4.6& 185 & $0.67 \pm 0.12$ & $7.93 \pm 0.07$ & 188 & $7.90 \pm 0.07$ & $7.97 \pm 0.12$ \\
6b..&                & 00:37 -- 01:36 &     &  59 & $0.66 \pm 0.09$ & $7.92 \pm 0.06$ &  59 & $7.99 \pm 0.09$ & $8.09 \pm 0.10$ \\
6c..&                & 02:14 -- 03:13 &     &     & indeterminate   &                 & 118 & $8.04 \pm 0.13$ & $8.09 \pm 0.14$ \\
7..& 2002 Jul 29     & 10:50 -- 11:26 & M4.7&   2 & $0.64 \pm 0.09$ & $7.91 \pm 0.05$ &  79 & $7.89 \pm 0.05$ & $7.91 \pm 0.07$ \\
8..& 2002 Oct 4      & 05:41 -- 05:56 & M4.0&  33 & indeterminate   &                 &  26 & $7.91 \pm 0.05$ & $8.03 \pm 0.12$ \\
9..& 2002 Dec 2      & 19:23 -- 19:32 & C9.6&  23 & $0.73 \pm 0.23$ & $7.96 \pm 0.12$ &  22 & $8.00 \pm 0.03$ & $8.05 \pm 0.10$ \\
10..& 2002 Dec 17/18 & 23:35 -- 01:01 & M1.6&  71 & $0.43 \pm 0.25$ & $7.73 \pm 0.20$ &  29 & $7.84 \pm 0.15$ & indeterminate \\
11..& 2003 Apr 23    & 01:00 -- 01:40 & M5.2&  43 & $0.73 \pm 0.27$ & $7.96 \pm 0.14$ & 101 & $7.99 \pm 0.05$ & $8.07 \pm 0.09$ \\
12..& 2003 May 29    & 01:10 -- 01:42 & X1.1&     & indeterminate   &                 & 118 & $7.90 \pm 0.05$ & $7.99 \pm 0.09$ \\
13..& 2003 Aug 19    & 10:00 -- 10:26 & M2.7& 121 & $0.85 \pm 0.37$ & $8.03 \pm 0.16$ & 106 & $7.94 \pm 0.05$ & $7.88 \pm 0.12$ \\
14..& 2003 Oct 22    & 20:16 -- 20:37 & M9.9&     &  indeterminate  &                 &  23 & $7.74 \pm 0.04$ & $7.83 \pm 0.08$ \\
15..& 2003 Oct 23    & 20:06 -- 20:37 & X1.1&  37 & $0.58 \pm 0.14$ & $7.86 \pm 0.10$ &  39 & $7.94 \pm 0.04$ & $8.06 \pm 0.09$ \\
16..& 2003 Nov  2    & 18:37 -- 18:59 & X8.3&  31 & $0.50 \pm 0.06$ & $7.80 \pm 0.05$ &  71 & $7.88 \pm 0.08$ & $8.03 \pm 0.12$ \\
17..& 2003 Nov  11   & 15:34 -- 16:33 & C8.5&  21 & $0.64 \pm 0.17$ & $7.91 \pm 0.10$ &  85 & $8.00 \pm 0.04$ & $8.13 \pm 0.09$ \\
18..& 2004 Jan  5    & 04:05 -- 04:52 & M6.9&  14 & $0.53 \pm 0.05$ & $7.82 \pm 0.04$ &  37 & $7.93 \pm 0.03$ & $8.08 \pm 0.02$ \\
19..& 2004 Jul  20   & 12:40 -- 13:41 & M8.7&  41 & $0.60 \pm 0.15$ & $7.88 \pm 0.10$ &  91 & $7.92 \pm 0.06$ & $8.00 \pm 0.10$ \\
20..& 2005 Jan  16   & 01:29 -- 02:27 & X2.6&  98 & $0.28 \pm 0.06$ & $7.55 \pm 0.08$ & 102 & $7.72 \pm 0.09$ & $7.90 \pm 0.18$ \\

\enddata
\tablenotetext{a} {$N_{\rm full}$ is the number of spectra analyzed with {\sc chianti$\_$full},  $N_{\rm cont}$ the number with {\sc chianti$\_$cont}.}

\tablenotetext{b} {``Indeterminate": standard deviation in $F> F$  or the standard deviation in $A({\rm Fe}) > 0.3$. See text for definition of $F$. }

%\tablenotetext{NOTE:--} {Mean value of $F$ from {\sc chianti$\_$full} fitting function weighted by standard deviation is $0.55 \pm 0.02$, corresponding to a mean value  $A({\rm Fe}) = 7.84 \pm 0.02$. Mean value of Fe abundance from the {\sc chianti$\_$cont} fitting function is $7.92 \pm 0.01$ (Fe line complex) and $8.08 \pm 0.02$ (Fe/Ni line complex).  }

\end{deluxetable}

% Subsection 2.3
\subsection{Spectral Analysis}

\begin{figure}
\begin{center}
\includegraphics[width=12cm,angle=0]{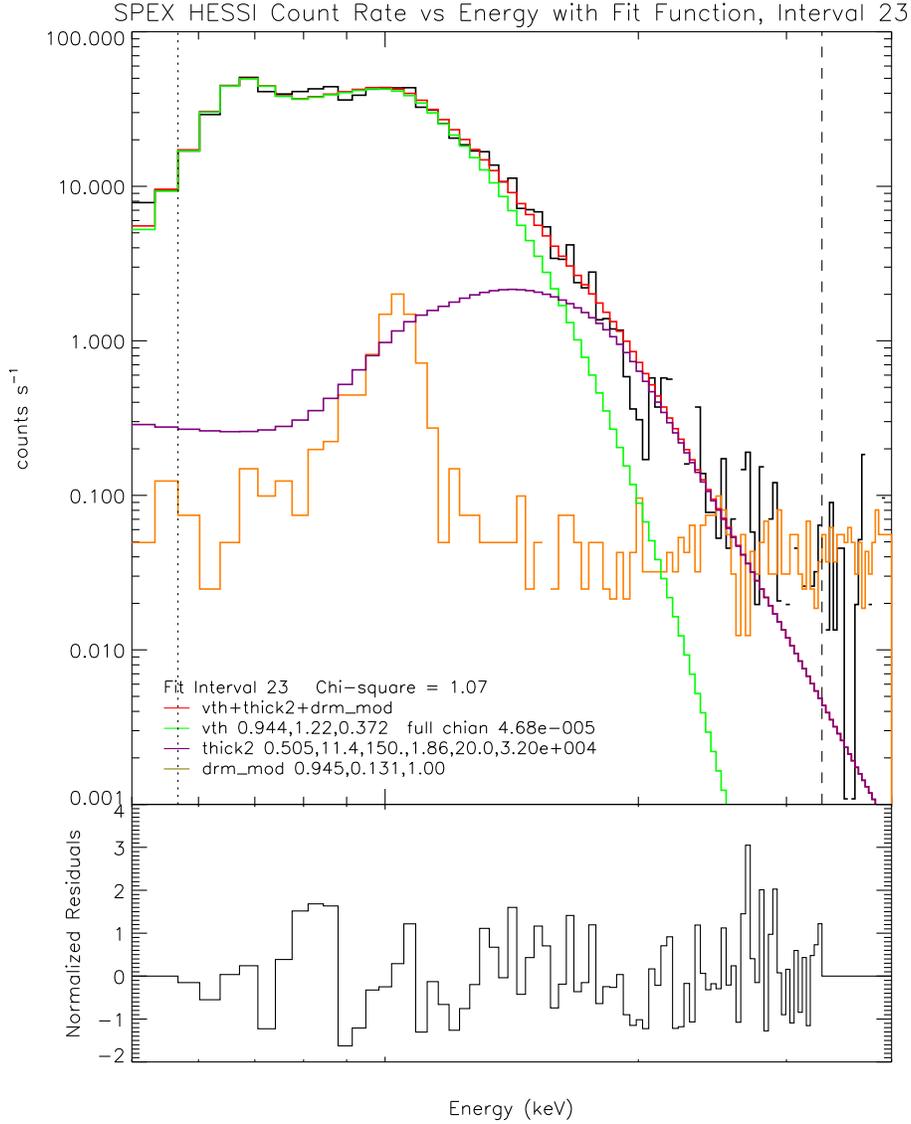}
\caption{(Upper panel) {\it RHESSI} detector~4 A1 background-subtracted count-rate spectrum (black histogram) for an interval during the decay of the 2003 October 23 flare (energy range is 5 to 40~keV). The least-squares fitted spectrum (red) consists of two components folded through the DRM: (a) a thermal spectrum (vth, in green) including all lines and continua in {\sc chianti}; (b) a nonthermal photon spectrum (thick2, dark purple). The function (drm$\_$mod)  allows the energy resolution and gain to vary to give the best fit to the Fe line complex. The energy range of the fit was 5.7--33~keV (vertical dotted and dashed lines). The reduced chi-squared $\chi_{\rm red}^2 = 1.07$. The pink histogram is the background spectrum. (Lower panel) Residuals normalized to the statistical $1 \sigma$ uncertainties in each energy bin.  } \label{count_rate_resids_CHIANTI_full}
\end{center}
\end{figure}

\begin{figure}
\begin{center}
\includegraphics[width=14cm,angle=0]{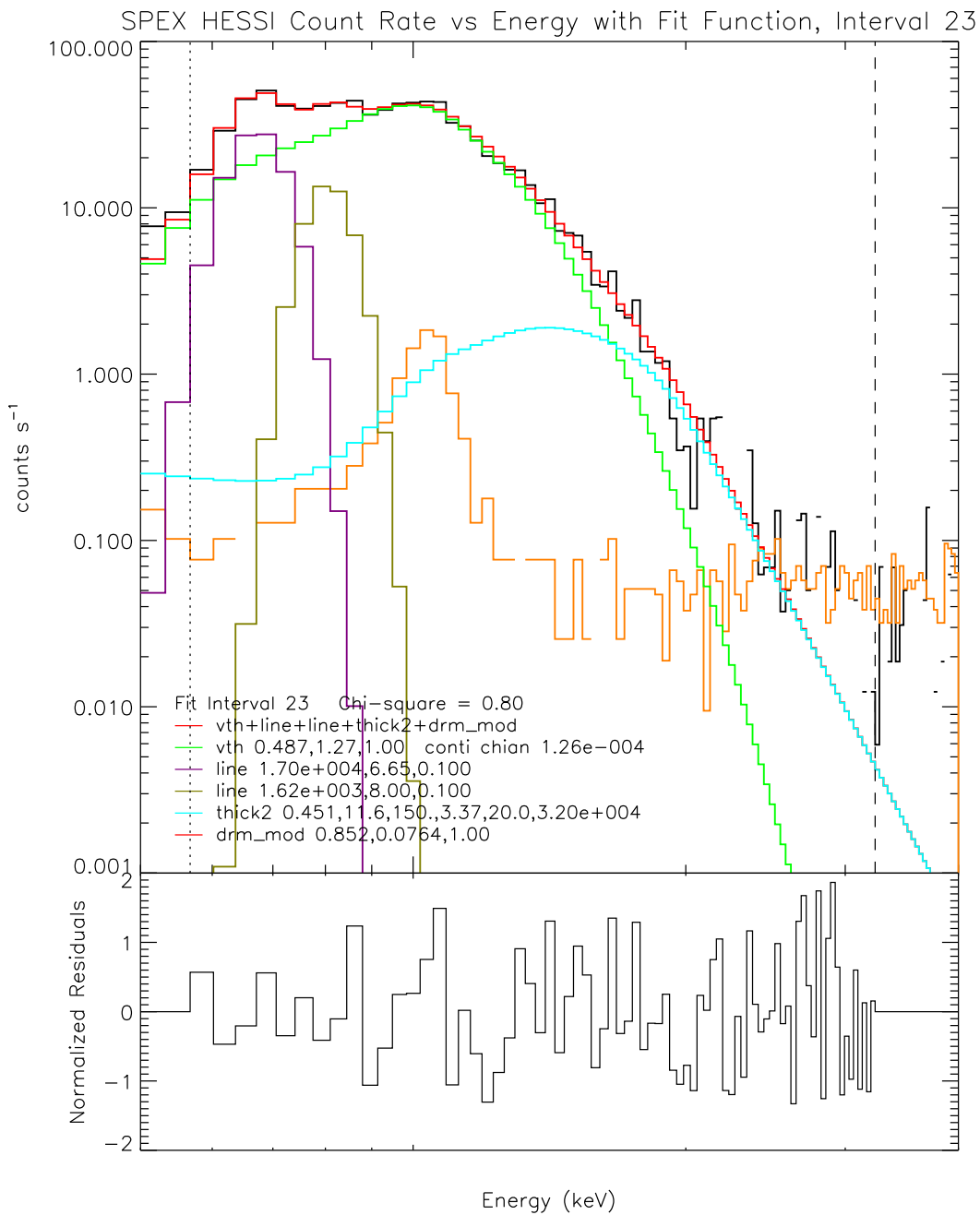}
\caption{Same as Figure~\ref{count_rate_resids_CHIANTI_full} with the thermal spectrum used for that plot replaced with the thermal continuum given by {\sc chianti} with coronal \citep{fel00} abundances and two lines with gaussian profiles at the mean energies of the Fe and Fe/Ni line complexes (6.65 and 8.0~keV respectively). The energy range is 5 to 40~keV. The value of $\chi_{\rm red}^2 = 0.80$. The mismatch in the residuals at $\sim 8$~keV has been significantly reduced. } \label{count_rate_resids_CHIANTI_cont}
\end{center}
\end{figure}

For each time interval, data and detector response matrix (DRM) files were extracted and read by OSPEX (Object Spectral Executive), an object-oriented IDL program recently substantially modified. The non-solar background spectrum was determined from the night-time parts of the {\it RHESSI} orbit. An isothermal fitting function was chosen to model the continuum and line emission, with goodness of fit determined by the reduced chi-squared $\chi_{\rm red}^2$, calculated as the weighted sum of the squares of the differences between background-subtracted count rates in each energy bin and the predicted count rates computed by folding the assumed incident photon spectrum through the DRM. The weights were determined from the predicted rates assuming Poisson statistics with zero systematic uncertainties. A graphical user interface allowed least-squares spectra with normalized residuals to be viewed (see  Figures~\ref{count_rate_resids_CHIANTI_full} and \ref{count_rate_resids_CHIANTI_cont}).  One form of the fitting function we chose, unavailable previously, is a thermal spectrum (vth) calculated from {\sc chianti}, including all lines and free--free and free--bound continua. The abundances of individual elements, most especially Fe, may be varied independently by a factor $F$ from a particular set of abundances, chosen in our case to be the ``coronal" set of \cite{fel92}. The ion fractions of \cite{bry09} were used. This value and $T_e$ and emission measure ($N_e^2 V$) were set as free parameters to be determined, as well as those describing any nonthermal continuum present. A further component of the fitting function (drm$\_$mod) allows for small modifications (gain and energy resolution) in the DRM. For this model fitting function ({\sc chianti$\_$full}), there is a total of 8 free parameters. There were slight disagreements in the fits, particularly around the energy of the Fe/Ni line complex ($\sim 8$~keV), with the {\sc chianti} spectrum underestimating the line flux, worsening the $\chi_{\rm red}^2$. Figure~\ref{count_rate_resids_CHIANTI_full} shows a {\it RHESSI} detector~4 A1 spectrum and fit with the {\sc chianti$\_$full} function and normalized residuals. For this fit, 80 energy bins were used with 8 free parameters defining the model spectrum, giving  $\chi_{\rm red}^2 = 1.07$. With 72 degrees of freedom, this indicates a probability of 32\% of exceeding this value through random statistical fluctuations in the count rates. The mismatch at $\sim 8$~keV is unlikely to be due to the omission in {\sc chianti} of $n>5$ lines of \ion{Fe}{25} \citep{phi08}; more probably it is due to an instrumental line from the tungsten collimator grids not allowed for in the DRM. The fit gives an abundance factor $F=0.372$, or $A({\rm Fe}) =7.67$.

% Fig. 1

A more general fitting function was also chosen that includes free--free and free--bound continua alone as calculated by {\sc chianti} and line features with Gaussian profiles to fit the Fe and Fe/Ni line complexes. This fitting function ({\sc chianti$\_$cont}) has a single temperature and emission measure describing the continuum, and the two line fluxes as free parameters. The line energies were kept fixed at their theoretical values (6.65~keV and 8~keV); there is only a slight temperature dependence of the line energies $\sim 0.05$~keV). As free--bound emission contributes to the continuum, there is some dependence on element abundances; as an initial estimate, we chose the \cite{fel92} abundance set for this calculation. As with the {\sc chianti$\_$full} function, gain and energy resolution adjustments with drm$\_$mod and a nonthermal continuum were included. This gave a total of 9 free parameters. The observed and fitted spectra for the same interval and detector are shown in Figure~\ref{count_rate_resids_CHIANTI_cont}, with normalized residuals. The value of  $\chi_{\rm red}^2 = 0.80$  implies a 89\% probability of exceeding this value; there is a fairly random distribution of residuals with energy over the fitted range (5.67--33~keV). Allowing the 8-keV line flux to be a free parameter removes the enhanced residuals at this energy.

As the Fe abundance will be derived from the line fluxes and the continuum values of $T_e$ and $N_e^2 V$, the dependence on element abundances of the continuum flux in the neighborhood of the Fe and Fe/Ni line complexes should be investigated, in particular the Fe abundance. Anticipating the discussion in Section~\ref{Fe_abund_anal}, an Fe abundance of $A({\rm Fe}) = 7.91$ or Fe/H $=8.13\times 10^{-5}$ is derived, i.e. a factor 1.55 less than the \cite{fel92} coronal value. The effect on the total continuum of this abundance difference can be partly tested with {\em RHESSI} software, since a provision is made in the analysis software to adjust the abundances of Si, S, Ca, Fe, and Ni by different factors in the continuum function. We took spectra during the 2003 October~23 flare with the {\sc chianti$\_$cont} model function but reducing {\em all} the Si, S, Ca, Fe, and Ni abundances by a factor 1.55. The abundances for these low-FIP elements would then be equal to the ``hybrid" abundances by \cite{flu99}. (There is nevertheless evidence that a constant reduction of all low-FIP elements is not observed; estimates of the flare potassium abundance from the RESIK instrument \citep{syl10a} indicate $A({\rm K}) = 5.86 \pm 0.23$ may be {\em enhanced} over the coronal value, 5.67, of \cite{fel92}.)  For the interval shown in Figure~\ref{count_rate_resids_CHIANTI_cont}, the temperature was practically unchanged at 1.28~keV = 14.8~MK but the emission measure was 16\% higher ($0.57 \times 10^{49}$~cm$^{-3}$). This difference folds directly into the derived Fe abundance.

The contribution that Fe alone makes to the free--bound and total continuum is, however, much less than the total of Si, S, Ca, Fe, and Ni. The exact contribution was calculated at temperatures typical of those found in this analysis, the results being given in Table~\ref{Fe_contr_cont}, where the percentage contribution of the Fe free--bound continuum at an energy of 10~keV (near the peak of the {\em RHESSI} count rate spectrum in the A1 state) is given to the total as a function of temperature (the contributions are similar at other energies). Thus, for $T_e=25$~MK, where the differences are most marked, the total continuum flux for a coronal Fe abundance ($A({\rm Fe}) = 8.10$) and an Fe abundance $A({\rm Fe}) = 7.91$ decreases by 6\%, from 15~\% to 21~\%. This means that the derived emission measure will increase by 6\% if the other elements (Si, S, Ca, and Ni) remain at their coronal \citep{fel92} abundances. At present, there is no recent detailed abundance analysis for these elements to confirm whether their abundances are the same as the \cite{fel92} abundances. However, it is worth pointing out that RESIK measurements of the continuum flux at somewhat lower energies (2.9--3.9~keV) \citep{phi10} are better described by coronal abundances \citep{fel92} than other abundance sets. Column~2 of Table~\ref{Fe_contr_cont} gives the percentage of the total free--bound continuum to the total continuum (free--free and free--bound) at 10~keV. The calculations in Table~\ref{Fe_contr_cont} are confirmed by more detailed calculations involving all significant elements (J. Sylwester, work in preparation).

% Table 3
\begin{deluxetable}{ccccc}
\tabletypesize{\scriptsize} \tablecaption{P{\sc ercentage of } T{\sc otal} C{\sc ontinuum at} 10 keV {\sc due to} Fe F{\sc ree--bound} R{\sc adiation} \label{Fe_contr_cont}}
 \tablewidth{0pt}
\tablehead{\colhead{Temperature (MK)} & \colhead{Ratio free--bound to} & \colhead{Photospheric} & \colhead{$2.6 \times$ phot. abund.} & \colhead{Coronal abundance} \\
& \colhead{total continuum (\%)$^a$} & \colhead{abundance (\%)} & \colhead{(\%)}  & \colhead{($4 \times$ photosph.) (\%)}}

\startdata
10 & 77 & 4 &  5 &  8 \\
12 & 73 & 6 & 8 & 12 \\
15 & 67 & 8 & 12 & 18 \\
20 & 58 & 9 & 15 & 21 \\
25 & 50 & 8 & 15 & 21 \\

\enddata
\tablenotetext{a}{Estimated with coronal \citep{fel00} abundances (for which the Fe abundance is 4 times photospheric.) }
\end{deluxetable}

% Fig. 2

\begin{figure}
\begin{center}
\includegraphics[width=16cm,angle=0]{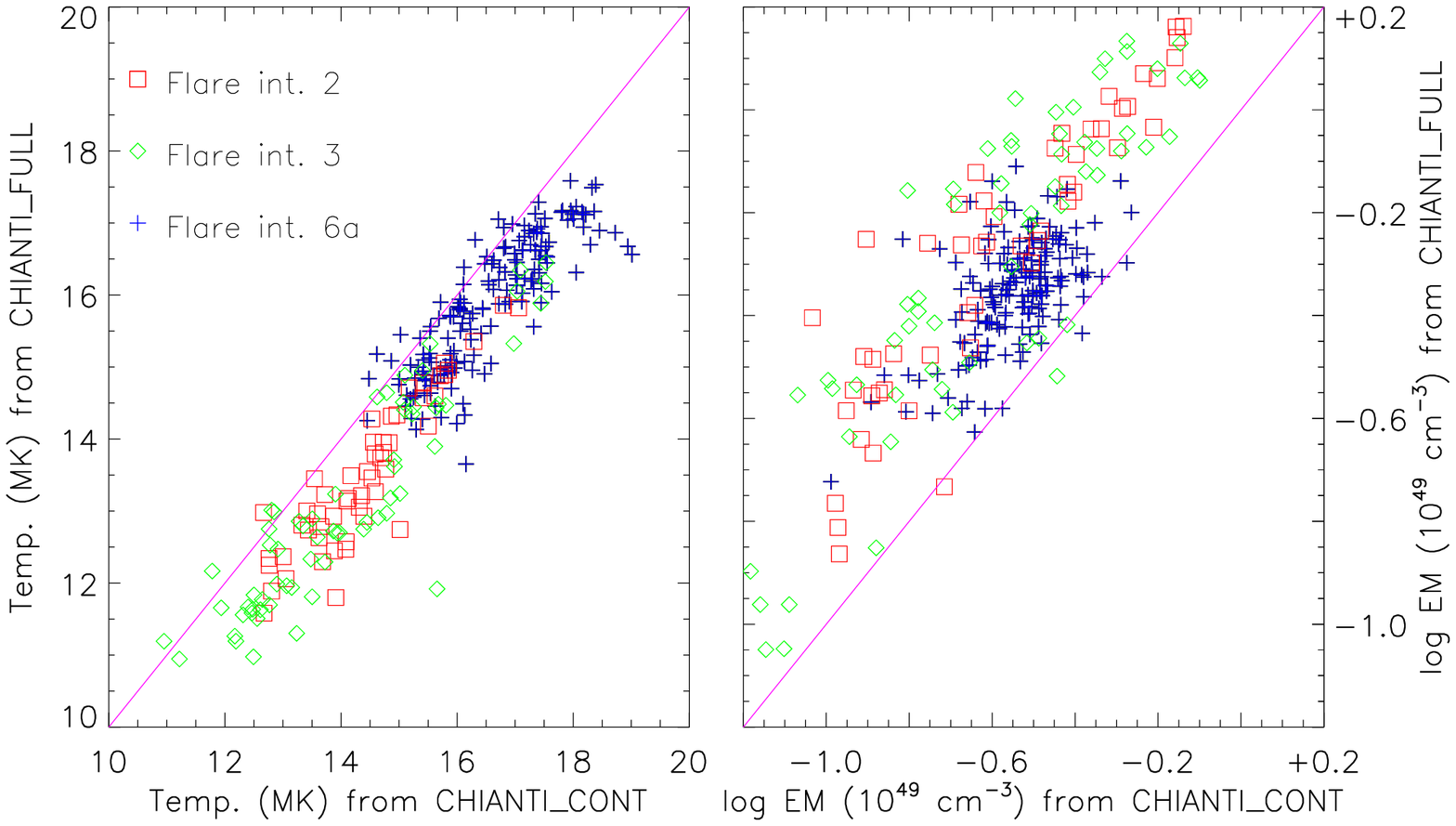}
\caption{Temperatures (MK) and logarithms of  emission measures (unit = $10^{49}$~cm$^{-3}$) from spectral fits to 3 flares using fitting function {\sc chianti$\_$full} plotted against those from fitting function {\sc chianti$\_$cont}. The three flare intervals are numbers 2, 3, and 6a in Table~\ref{anal_ints}, with symbols indicated in the legend. } \label{compare_T_EM}
\end{center}
\end{figure}

Because of the mismatch using the {\sc chianti$\_$full} fitting function at $\sim 8$~keV, there were fewer spectral fits having small values of reduced $\chi_{\rm red}^2$ than with the {\sc chianti$\_$cont} fitting function. Temperatures and emission measures from {\sc chianti$\_$cont} and {\sc chianti$\_$full} are compared in the plot shown in Figure~\ref{compare_T_EM} for time intervals during three flares for which $\chi_{\rm red}^2 < 1.5$. It is evident that temperatures from the {\sc chianti$\_$full} model are smaller than those from {\sc chianti$\_$cont} by $\sim 1$~MK and emission measures larger by a factor $\sim 2.5$ (0.4 in the logarithm). Although the specific reasons for these differences are unclear, they may be due to attempts in the fitting process with {\sc chianti$\_$full} to correct for the $\sim 8$~keV mismatch with a continuum function that is slightly too steep at energies $\gtrsim 8$~keV.

% Fig. 3

In our previous analysis, we compared spectral fits from seven of the nine {\em RHESSI} detectors suitable for spectral analysis in the low-energy region (detectors 2 and 7 have poorer energy resolution and higher energy threshold and so they were not used). Here, this comparison was done more systematically using a 30-minute time interval in the decline of the flare on 2002 July~26 with the {\sc chianti$\_$cont} fitting function. From a detailed analysis of the  temperature and emission measure over this period, the mean $1\sigma$ uncertainties (in MK) for the detectors 1, 3, 4, 5, 6, 8 and 9 were found to be respectively 1.20, 1.19, 1.05, 3.02, 1.12, 1.08, 1.16. They are thus smallest for detector 4 and largest for detector 5. This is also true for the uncertainties in the emission measure estimates. In light of these results, our choice of spectral fits from detector~4 in our earlier work appears to be vindicated, and so we chose to fit spectra from this detector.

% Subsection 2.4
\subsection{Use and Validation of CHIANTI Spectra}

The {\em RHESSI} analysis code now uses version~6 of the {\sc chianti} atomic code and database \citep{der09} which includes the best atomic data available for the lines in the {\em RHESSI} range discussed here. In particular, an abundance of Fe is directly determined from comparison of the  Fe line and Fe/Ni line complex fluxes with that from {\sc chianti} using $T_e$ and emission measure from the continuum rather than with our earlier work \citep{phi06} in which the equivalent width of each line complex was determined from {\em RHESSI} spectra and compared with calculations based on the sum of individual lines in each complex by \cite{phi04}. Expressed as contribution functions $G_{\rm Fe}(T_e)$ and $G_{\rm Fe/Ni}(T_e)$, or the amount of emission from the Fe line complex (defined to be all lines in the 6.391--7.005~keV range) and Fe/Ni line complex (all lines in the 7.728--8.907~keV range) from a plasma with unit volume emission  measure, there are differences of only a few per cent between the \cite{phi04} and {\sc chianti} v.~6 calculations for the Fe line. There are more significant differences for the Fe/Ni line for $T_e \lesssim 12$~MK, for which the {\sc chianti} v.~6 calculations are higher by amounts that depend on $T_e$. These are due to the addition of \ion{Fe}{24} satellites near the \ion{Fe}{25} $1s^2 - 1snp$ ($n=4$, 5) lines not included by \cite{phi04} or earlier versions of {\sc chianti}. The {\sc chianti} v.~6 calculations for $G_{\rm Fe/Ni}(T_e)$ are still incomplete in that \ion{Fe}{25} $1s^2 - 1snp$ ($n > 5$) lines and the associated satellites are not included. However, based on work by \cite{phi08}, this is unlikely to lead to an underestimate of $G_{\rm Fe/Ni}(T_e)$ by more than a few percent in the range of temperatures (approximately 10--22~MK) considered in this work.

Validation of {\sc chianti} spectra is possible by comparing with solar flare spectra from the {\em Solar Maximum Mission} Bent Crystal Spectrometer for the Fe line complex. There are small differences apparent in the ratio of some of the intense \ion{Fe}{24} satellites to the \ion{Fe}{25} resonance line which provides the means of determining $T_e$ in high-resolution, crystal spectrometer data. The BCS spectra analyzed by \cite{lem84} give $T_e = 15.0 \pm 0.5$~MK, whereas a re-analysis with {\sc chianti} v.~6 leads to a higher temperature, 16.5~MK. The difference is probably due to improved atomic data for both the \ion{Fe}{24} satellites and the collisional rates for the \ion{Fe}{25} resonance line. Rather large differences (up to 0.01~keV) are apparent in the energies of \ion{Fe}{21}--\ion{Fe}{23} satellites, though these are not likely to affect the total flux of the Fe line complex. There are no high-resolution flare spectra of the Fe/Ni line complex, so validation is not possible.

% Section 3
\section{Fe ABUNDANCE ANALYSIS}\label{Fe_abund_anal}

Applying the {\sc chianti$\_$full} fitting function to {\em RHESSI} spectra gives temperature, emission measure, and the abundance factor $F$, which is the factor applied to the baseline Fe and Ni abundance and is determined by the fluxes of the Fe and Fe/Ni line complexes. The baseline abundance set used was the coronal one of \cite{fel00} (for Fe this is $A({\rm Fe})_{FL} = 8.10$). For all the flares analyzed, the values of $F$ are practically constant with time and are unrelated to variations in $T_e$. Since the Fe and Fe/Ni line features in the energy range chosen are nearly entirely due to iron, the measured Fe abundance is $F$ multiplied by the coronal \citep{fel00} value, or in logarithmic terms, $A({\rm Fe})_{\rm meas} = {\rm log}_{10}F + A({\rm Fe})_{FL}$. Table~\ref{anal_ints} gives the mean value of $F$ and standard deviation for each of the time periods listed with the number of spectra used to derive $F$ having $\chi_{\rm red}^2 < 1.5$. For a few flares there were too few good-quality spectra to give a reliable value of $F$ (marked ``indeterminate"). For the remaining 18 sets of spectra, the mean value of $F$ is $0.55 \pm 0.08$ (s.d.), and so $A({\rm Fe})_{\rm meas} = 7.90 \pm 0.02$. This is several standard deviations less than $A({\rm Fe})_{FL}$, despite the fact that the value of $F$ is poorly determined for some flares. It is a factor 2.5 more than the photospheric Fe abundance  of \cite{asp09}. To determine the abundance of Fe from the {\sc chianti$\_$cont} model, we followed the procedure for analyzing spectra from the RESIK crystal spectrometer \citep{syl10a,syl10b}. For a spectrum in a particular ($i$th) {\em RHESSI} time interval, the iron abundance is determined from the flux $\mathfrak{F}_i$ of either the Fe line or the Fe/Ni line complexes using

% Equation 1
\begin{equation}
f_i ({\rm Fe}) = \frac{\mathfrak{F}_i}{G(T_i) EM_i}
\end{equation}\label{eq_for_f}

\noindent where the temperature $T_i$ and emission measure $EM_i$ are output from the $i$th best-fit model function. Thus, while $f_i$, like $F_i$ in the {\sc chianti$\_$full} fitting model, is a factor multiplying the \cite{fel00} Fe abundance to give the measured Fe abundance, it is defined in terms of the measured fluxes of either the Fe or Fe/Ni line complex, the continuum temperature $T_i$ and emission measure $EM_i$, and the contribution function $G_{\rm Fe} (T_i)$ or $G_{\rm Fe/Ni} (T_i)$.

\begin{figure}
\epsscale{1.}
\plottwo{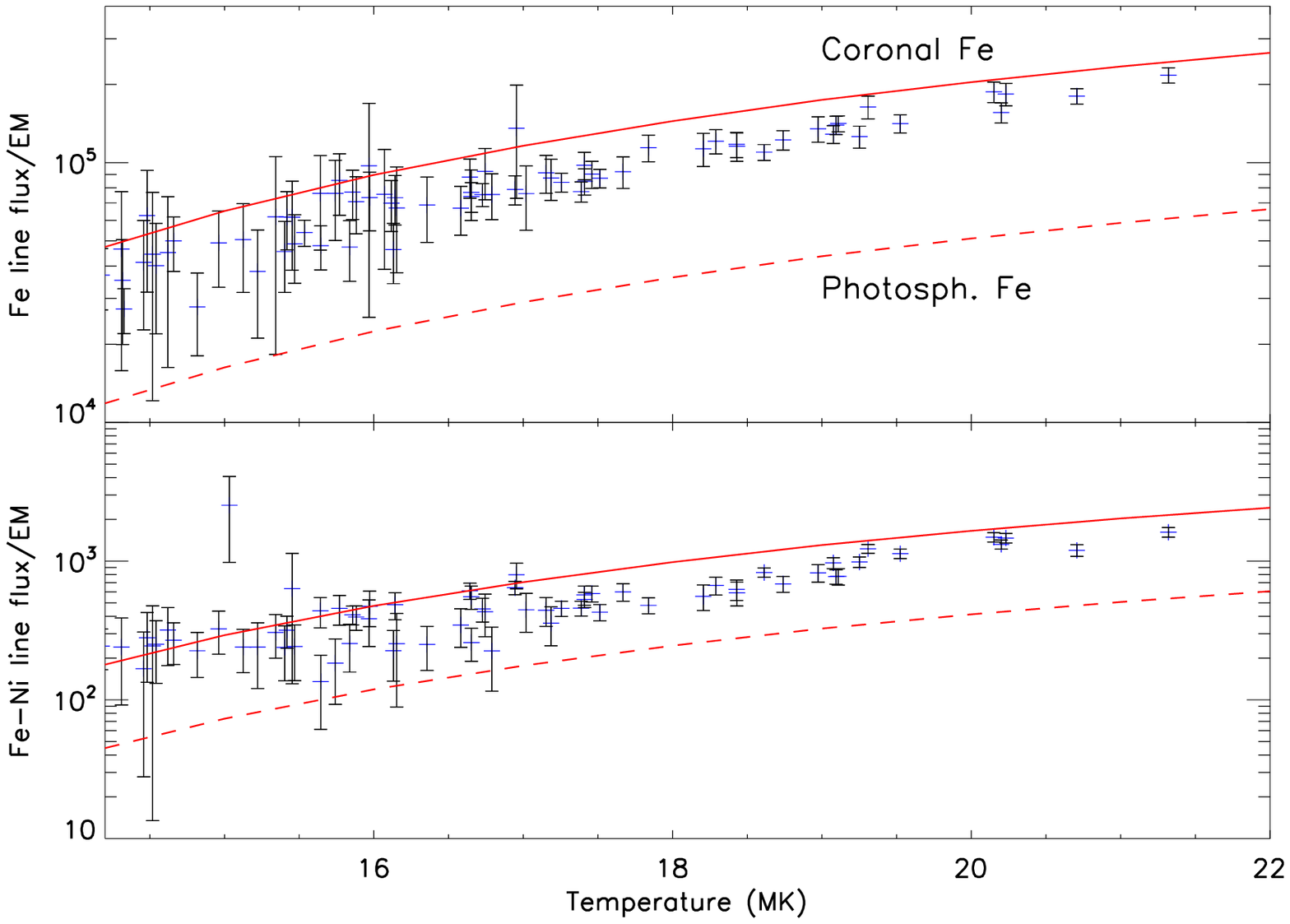}{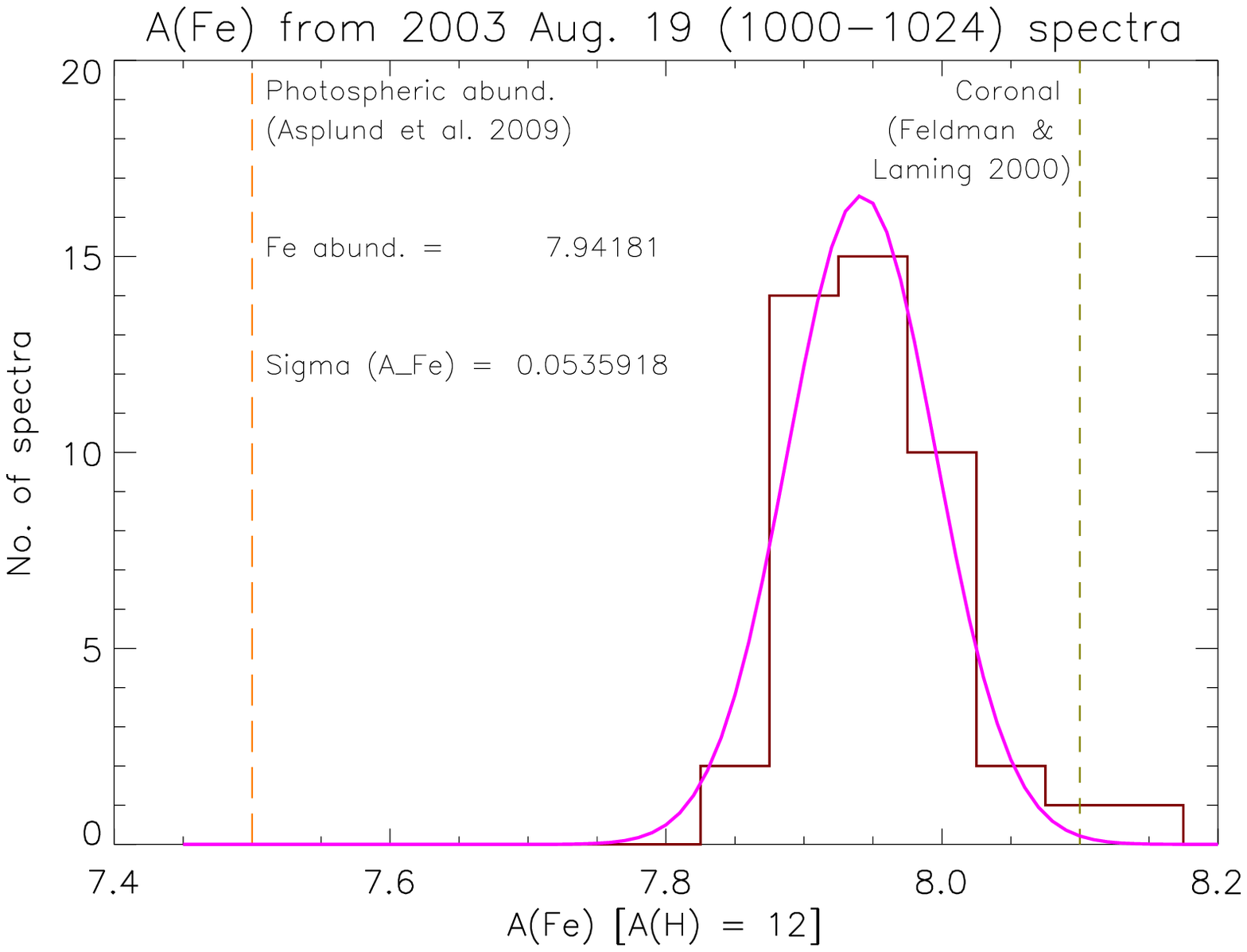}
\caption{(Left) Line flux/emission measure ($10^{49}$ cm$^{-3}$) for the Fe (6.65 keV) and Fe/Ni (8 keV) line complexes vs. $T_e$ derived from {\em RHESSI} A1 spectra during the decline of the flare on 2003 August~19. The $G_{\rm Fe}(T_e)$ and $G_{\rm Fe/Ni}(T_e)$ curves calculated from {\sc chianti} are shown for coronal and photospheric Fe abundances. (Right) Distribution of the Fe abundances derived from the Fe line fluxes (histogram), with coronal and photospheric Fe abundances shown. The best-fit gaussian (smooth curve) indicates a mean value  $A({\rm Fe}) = 7.94 \pm 0.05$. \label{GofT_20030819}}
\end{figure}

The estimated fluxes in the Fe line and Fe/Ni line complexes divided by the continuum emission measure when plotted against continuum temperature can be compared with the theoretical  $G_{\rm Fe}(T_e)$ and $G_{\rm Fe/Ni}(T_e)$ functions calculated with $A({\rm Fe})_{FL} = 8.10$ and the photospheric abundance ($A({\rm Fe}) = 7.50$). This is done in Figure~\ref{GofT_20030819} (left) for the 2003 August~19 flare. The uncertainties (from the OSPEX software) are larger for the Fe/Ni line complex, which for the temperatures analyzed here is about a factor 10--20 less intense than the Fe line complex, and are larger for lower-temperature spectra late in the flare when the emission was weaker.

% Fig. 4

\begin{figure}
\epsscale{.80}
\plotone{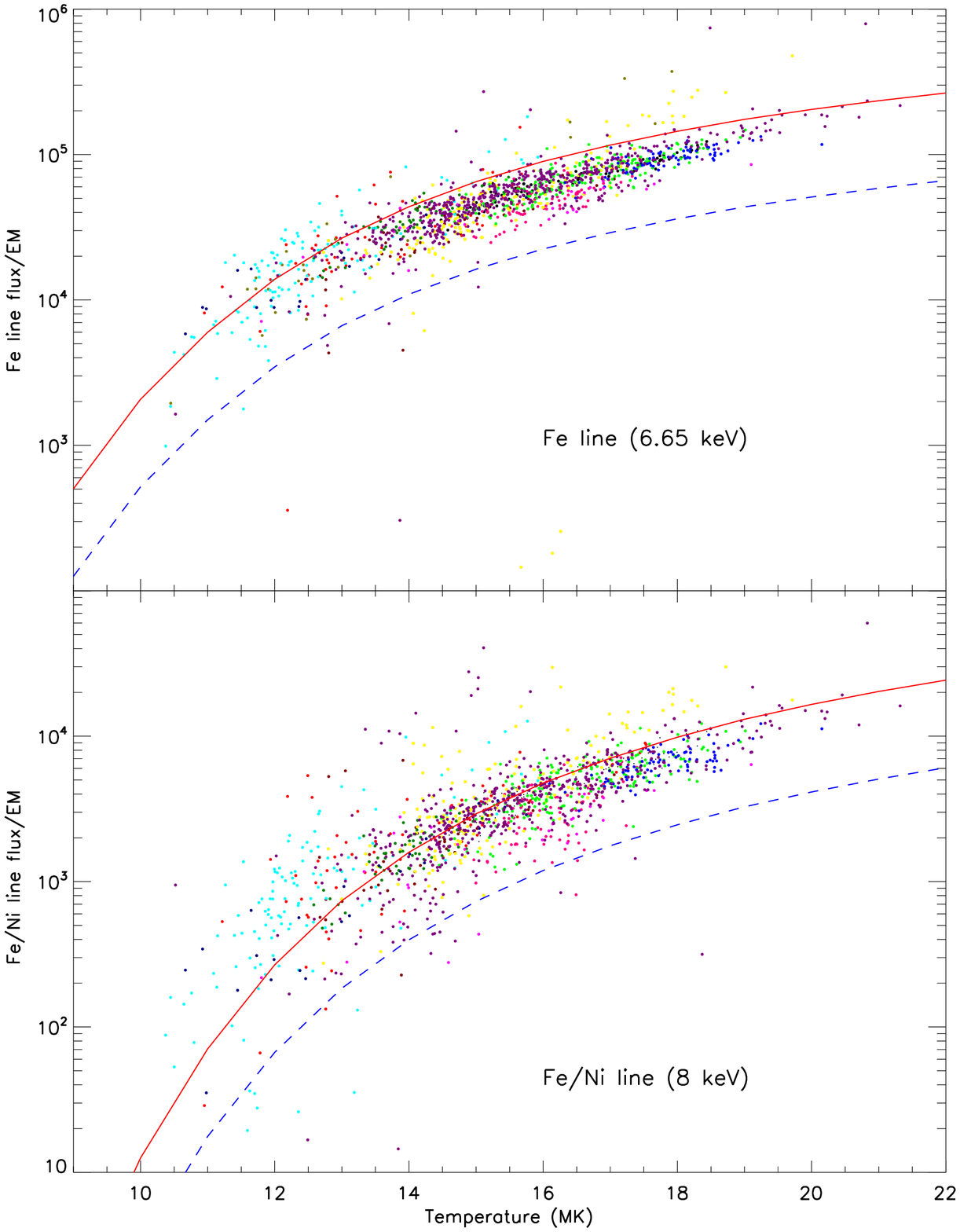}
\caption{Line flux/emission measure ($10^{49}$ cm$^{-3}$) plotted against $T_e$ for all flares in this analysis for the Fe (6.65 keV) (upper panel) and Fe/Ni (8 keV) (lower panel) line complexes derived from {\em RHESSI} A1 spectra. The calculated $G_{\rm Fe}(T_e)$ and $G_{\rm Fe/Ni}(T_e)$ curves are shown as solid lines for coronal \citep{fel00} and dashed lines for photospheric \citep{asp09} Fe abundances.  \label{line_flux_allflares}}
\end{figure}

Values of measured Fe abundances from Eq.~(1) were obtained for each spectral interval in this and other flares, and histogram distributions found for the Fe and Fe/Ni line complexes. This is illustrated for the Fe line complex in Figure~\ref{GofT_20030819} (right) where numbers of values falling into intervals of 0.05 in $A({\rm Fe})$ are plotted, together with a best-fit gaussian curve. From the best-fit curves, a mean value of $A({\rm Fe})$ and standard deviation can be derived. The values derived are listed for each of the time periods in the last two columns of Table~\ref{anal_ints}. For some time periods, the Fe/Ni line complex was too weak to be measured so the Fe abundance could not be derived; the abundances are marked ``indeterminate" when the standard deviation in $A({\rm Fe})$ was larger than 0.3 (corresponding to a factor 2 in the abundance estimate). Similar plots to Figures~\ref{GofT_20030819}  were done for the time ranges in Table~\ref{anal_ints}. The total of all flares for the Fe line and Fe/Ni line complexes is shown in Figure~\ref{line_flux_allflares}. There is a clear trend for the values of flux divided by emission measure to be below the coronal $G(T_e)$ curves by a constant factor for the Fe line complex, suggesting (as with the {\sc chianti$\_$full} analysis) that the Fe abundance is smaller than the \cite{fel00} value, but larger than the photospheric value. For the Fe/Ni line complex, the trend is similar at high temperatures ($T_e \gtrsim 16.5$~MK, but the points become steadily higher than the coronal abundance curve for decreasing temperatures. This may be because of the instrumental line at $\sim 8$~keV mentioned earlier.

\begin{figure}
\epsscale{.80}
\plotone{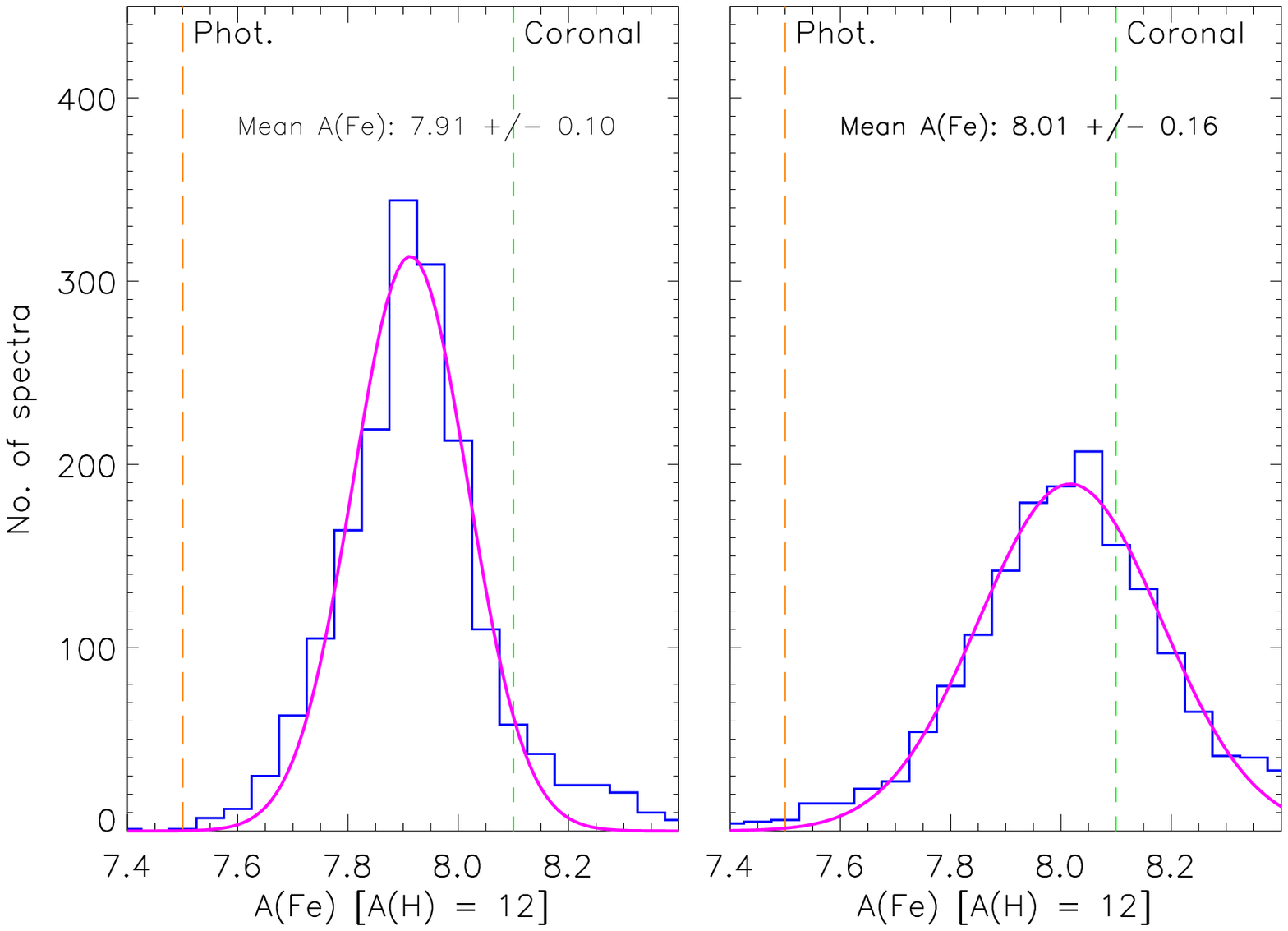}
\caption{Distribution of values of derived Fe abundance values, $A({\rm Fe})_i$, from fits to {\em RHESSI} spectra using {\sc chianti$\_$cont} for all flares in this analysis, compared with best-fit Gaussian curves. Left: from the Fe (6.65 keV) line complex; right: from the Fe/Ni (8 keV) line complex.  The coronal and photospheric Fe abundances are shown as dashed vertical lines. \label{Fe_abund_hist}}
\end{figure}

Figure~\ref{Fe_abund_hist} shows the distribution of Fe abundance values for spectra during all flares lumped together for the Fe line and Fe/Ni line complexes, with best-fit gaussian curves. The mean value of $A({\rm Fe})$ is $7.91 \pm 0.10$ from the Fe line, $8.01 \pm 0.16$ from the Fe/Ni line complex. The larger Fe abundance and larger uncertainty from the Fe/Ni line reflects the departure of the points from the theoretical $G_{\rm Fe/Ni}(T_e)$ curve, so the Fe abundance from the Fe line complex is clearly the preferred value. We note that the continuum in this analysis has a small  contribution from Fe emission, so there is a slight dependence on the coronal abundances used, which was the \cite{fel92} set. Use of an Fe abundance  $A({\rm Fe}) = 7.91$ instead of $A({\rm Fe})_{FL} = 8.10$, if the elements Si, S, Ca, and Ni remain at their coronal values, should lead to emission measures that are  $\sim 6$\% larger. Eq.~(1) indicates that the derived value of $f_i$ and therefore the Fe abundance using the {\sc chianti$\_$cont} emission model with coronal abundances for the continuum will be too large by 6\% (0.03 in the logarithm). The abundance variation from flare to flare, in spite of the factor-of-100 range in the {\em GOES} X-ray importance and {\em RHESSI} count rates, is very small and well within the standard deviations for each flare.

% Fig. 5

% Fig. 6

% Section 4
\section{Discussion and Conclusions}

Our analysis of {\em RHESSI} solar flare spectra gives estimates of the Fe abundance from two different emission models, {\sc chianti$\_$full} (continuum and lines as given by the {\sc chianti} atomic code) and {\sc chianti$\_$cont} (continuum given by {\sc chianti},  Fe and Fe/Ni line complexes at 6.65~keV and 8~keV separately fitted). Both methods give Fe abundance values that are constant to within the estimated uncertainties and higher than those derived from the photosphere  or meteorites. The {\sc chianti$\_$full} analysis was found to suffer from poor fits to the spectral region around the Fe/Ni line complex at 8~keV, leading to generally higher values of $\chi_{\rm red}^2$. Using only spectral fits with $\chi_{\rm red}^2 < 1.5$, we found from 18 time intervals that the mean $A({\rm Fe})$ to be $7.87 \pm 0.02$, the uncertainty being in the mean values for each of the 18 intervals. The {\sc chianti$\_$cont} analysis has the advantage that the Fe and Fe/Ni line complexes can be fitted separately and $A({\rm Fe})$ can be derived for each. The possible presence of an instrumental line at $\sim 8$~keV adds to the observed Fe/Ni line complex and so the measured flux  is an upper limit to the solar value. This probably explains the 23\% difference between the values $A({\rm Fe}) = 7.91 \pm 0.10$ for the Fe line during 25 time intervals and $A({\rm Fe}) = 8.01 \pm 0.16$ for the Fe/Ni line during 21 time periods, though the Fe/Ni line complex is also a much weaker feature. When all the {\sc chianti$\_$cont} observations are lumped together (Figures~\ref{line_flux_allflares} and \ref{Fe_abund_hist}), it is clear, particularly from the Fe line, that the Fe abundance is constant to within estimated uncertainties from flare to flare. The distribution of all the estimates from the Fe line leads to an Fe abundance given by $A({\rm Fe}) = 7.91 \pm 0.10$, which is our preferred value.

Our value is higher by a factor $2.6 \pm 0.6$ than Fe abundance estimates from the photosphere (e.g. $A({\rm Fe}) = 7.50 \pm 0.04$: \cite{asp09}) and by a factor $2.9 \pm 0.7$ than those from meteorites ($7.45 \pm 0.01$: \cite{lod09}).
However, it is lower than those given for coronal plasmas, which range from 7.65 \citep{par77,flu99} to 8.19 \citep{whi00}; it is a factor $1.55 \pm 0.5$ lower than that given by \cite{fel00} and \cite{fel92}, viz. $A({\rm Fe}) = 8.10$, which is used as the coronal abundance set in {\sc chianti}. The preliminary results for flares observed with the XRS instrument on Mercury MESSENGER \citep{den08} give an Fe abundance that is a factor 2.3 times photospheric \citep{gre98}, i.e. $A({\rm Fe}) = 7.9$, which is consistent with our preferred value. Our value is within $1\sigma$ of the Fe abundance of Table~2 of \cite{flu99}, $A({\rm Fe}) = 7.83$, for their ``hybrid" abundance model, taking account of the uncertainties in both our value and the hybrid model. Our value of $7.91 \pm 0.10$ is in very close agreement (0.02 less)  with the energetic particle abundances reported by \cite{rea95} for gradual events, viz. $A({\rm Fe}) = 7.93 \pm 0.01$, suggesting a relationship of the fractionation process involved in the formation of hot flare plasmas and the acceleration of solar energetic particles. The constancy of the {\em RHESSI} Fe abundance estimates points to fractionation processes in flares, at least their declining stages, to be similar to those for the quiet Sun. It is rather against expectations, as discussed by \cite{asp09}, who state that ``the degree of chemical separation varies significantly, being more severe in regions of higher solar activity." Though our measurements are during the decay of flares, the flare plasma is unlikely to be mixed with other plasma after the impulsive stage when (as is widely accepted) chromospheric evaporation occurs.

\begin{figure}
\epsscale{.80}
\includegraphics[width= 12cm,angle=0]{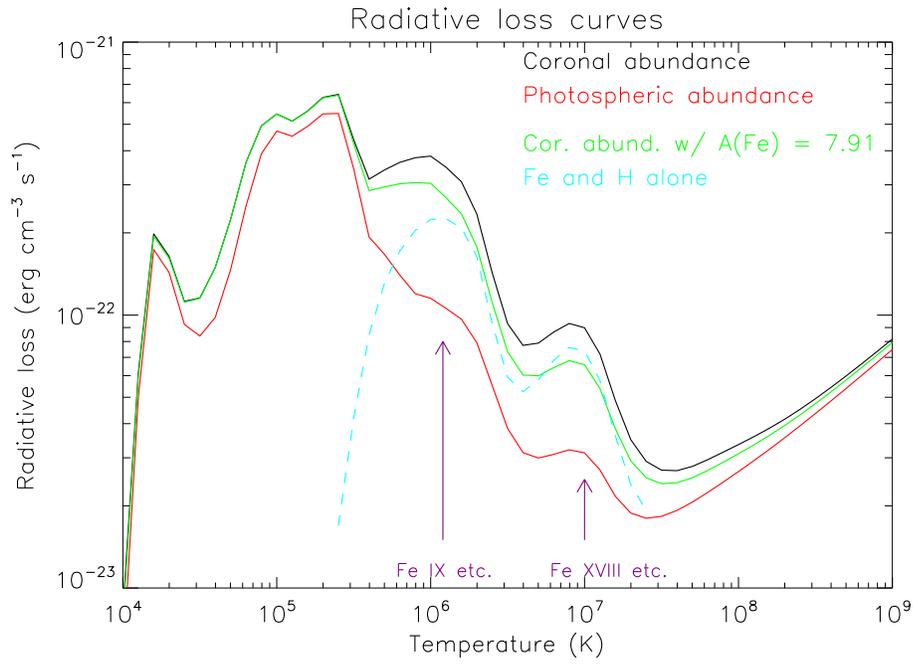}
\caption{Radiation loss curves for photospheric \citep{asp09} and coronal \citep{fel00} abundances, and coronal abundances with the Fe abundance from this work ($A({\rm Fe}) = 7.91$).  \label{rad_loss}}
\end{figure}

If the Fe abundance derived here for flares is representative of the quiet solar corona and active-region levels, there are consequences for the radiation loss curve (Figure~\ref{rad_loss}). Iron ions are an important emitter for solar plasmas in the temperature range spanning values of the quiet corona to the tens of MK of solar flares.  A maximum in radiation loss at $\sim 1$~MK is due to emission lines of \ion{Fe}{9}--\ion{Fe}{15} between 171~\AA\ and 284~\AA, and a second maximum at $\sim 10$~MK is due to \ion{Fe}{17}--\ion{Fe}{24} X-ray lines. The precise abundance of iron is therefore important for studies of the energy balance in quiet coronal or flare loops, particularly for higher densities when radiation loss may dominate conduction losses \citep{kli08}. If the value obtained here, $A({\rm Fe}) = 7.91$, applies to all coronal plasmas with temperatures 1--20~MK, the radiation loss curve will be correspondingly modified. Figure~\ref{rad_loss} shows the radiation loss for optically thin plasma with this temperature range and for Fe abundances with photospheric and coronal values including our own. The contribution from Fe ions and hydrogen alone is also shown, with maxima at $\sim 1$~MK and $\sim 10$~MK. (The maximum at $\sim 20000$~K is due to Ly-$\alpha$ emission of hydrogen.) The radiation loss is less for a photospheric Fe abundance than a coronal Fe abundance by a  factor $\sim 3.4$ for quiet coronal loops ($T_e \sim 1$~MK) and a factor $\sim 3.0$ for flare loops with $T_e \sim 10$~MK. For $A({\rm Fe}) = 7.91$ obtained in this work, the radiation loss is less than that for coronal Fe abundance \citep{fel92} by a factor 0.8 ($T_e \sim 1$~MK) and 0.72 ($T_e \sim 10$~MK). The radiation loss curve approximated by \cite{kli08} with a piece-wise continuous temperature-dependent form assumed abundances that are twice those of \cite{mey85}; as \cite{mey85} gives $A({\rm Fe}) = 7.6$ for the corona, the \cite{kli08} value (7.9) is very nearly the one obtained here.

% Fig. 7

The various models advanced up to the mid-1990s for explaining the FIP effect have been reviewed by \cite{hen98}. Fractionation according to the first ionization potential is generally explained by the fact that low-FIP (FIP $\lesssim 10$~eV) elements are partly ionized in the photosphere but high-FIP elements are predominantly neutral. It is difficult to obtain from any of these early models a quantitative enhancement of low-FIP elements in coronal plasmas, however, and it appears to be a feature of the present work that for flares with a range of X-ray importance that the enhancement of Fe is a particular value, constant from flare to flare. The constancy of the Fe abundance in flare plasmas is also consistent with the analysis of 2795 RESIK spectra indicating a constant enhancement (of a factor 5 over the photospheric abundance) of potassium in flares \citep{syl10a} and an argon abundance that is also constant (in this case to within 20\% of photospheric Ar abundance proxies \citep{syl10b}). The more recent FIP model of Laming \citep{lam04,lam09} involves the ponderomotive force that arises when Alfv\'en waves pass from the chromosphere to the corona, and definite predictions about the enhancements of various elements can be made according to the dimensions of the coronal loop that the waves are incident on, its magnetic field, and the wave intensity. A particular example is given for a loop with length of 100,000~km and field strength of 7~G. {\em RHESSI} flare loops are likely to be much shorter, but it is interesting that for a fairly wide range of wave energy fluxes the coronal enhancement of Fe is between 2 and 3, as obtained in this work.

In a wider context, it has been found that the occurrence of giant planets around stars depends sensitively on metallicity, specifically the Fe abundance \citep{fis05,val08}. This refers to Fe abundances in the stellar photospheres. However, it would be interesting to use the methods of this work to derive the coronal or flare abundance of Fe from the 6.65~keV line feature and compare with photospheric abundances to see if there are correlations of the difference with the probability of planet formation.

\acknowledgments
We thank A. K. Tolbert and R. A. Schwartz for their invaluable help in the data analysis and to A. Gopie for initial data reduction. J. Sylwester and B. Sylwester are thanked for the use of their methodology in the derivation of the Fe abundance in this work.  K.~J.~H.~P. acknowledges support from NASA through ADNET (under the SESAA-II contract, NNG06EB68C) for a visit to Goddard Space Flight Center in 2008 and support from a National Research Council Senior Research Associateship during the original conception of this work. {\sc chianti} is a collaborative project involving the US Naval Research Laboratory, the Universities of Florence (Italy) and Cambridge (UK), and George Mason University (USA). We are grateful to the authors of the {\sc chianti} code for continued help in adding data to the spectral regions discussed here.

\end{document}